\documentclass[12pt]{article}
\usepackage{newtxtext,newtxmath}
\usepackage{graphicx}
\usepackage[letterpaper,margin=1in]{geometry}
\renewenvironment{abstract}
	{\quotation}
	{\endquotation}

\date{}

\makeatletter
\renewcommand{\fnum@figure}{\textbf{Figure \thefigure}}
\renewcommand{\fnum@table}{\textbf{Table \thetable}}
\makeatother
\usepackage{scicite}
\usepackage{url}
\usepackage{hyperref}
\usepackage{xurl}
\newcommand{\added}[1]{#1}
\newcommand{\deleted}[1]{}
\newcommand{\replaced}[2]{#2}
\usepackage[super]{cite}

\usepackage[font={small,stretch=0.9},labelfont=bf]{caption}
\def\scititle{
	High-dimensional structure underlying individual differences in naturalistic visual experience
}
\title{\bfseries \boldmath \scititle}

\author{
	Chihye Han$^{1}$,
	Michael F. Bonner$^{1\ast}$\and
	\small$^{1}$Department of Cognitive Science, Johns Hopkins University, Baltimore, MD 21218 \and
	\small$^\ast$Corresponding author. Email: mfbonner@jhu.edu\and
}

\begin{document} 

\maketitle

\begin{abstract} \bfseries \boldmath
How do different brains create unique visual experiences from identical sensory input? While neural representations vary across individuals, the fundamental architecture underlying these differences remains poorly understood. Here, we reveal that individual visual experience emerges from a high-dimensional neural geometry across the visual cortical hierarchy. Using spectral decomposition of fMRI responses during naturalistic movie viewing, we find that idiosyncratic neural patterns persist across multiple orders of magnitude of latent dimensions. Remarkably, each dimensional range encodes qualitatively distinct aspects of individual processing, and this multidimensional neural geometry predicts subsequent behavioral differences in memory recall. These fine-grained patterns of inter-individual variability cannot be reduced to those detected by conventional intersubject correlation measures. Our findings demonstrate that subjective visual experience arises from information integrated across an expansive multidimensional manifold. This geometric framework offers a powerful new lens for understanding how diverse brains construct unique perceptual worlds from shared experiences.
\end{abstract}

\section{Introduction}
The human brain transforms visual input into subjective perceptual experiences. When presented with identical visual stimuli, individuals exhibit idiosyncratic patterns of neural activity across cortical regions involved in visual processing \cite{Charest_Kriegeskorte_2015}. Idiosyncratic neural activity patterns remain stable across time \cite{finn2015functional, gratton2018functional} and carry meaningful information about behavior  \cite{finn2015functional, Finn_Glerean_Khojandi_Nielson_Molfese_Handwerker_Bandettini_2020}. Individuals with shared personality traits \cite{Finn_Corlett_Chen_Bandettini_Constable_2018}, shared context \cite{yeshurun2017same}, or shared perspectives \cite{lahnakoski2014synchronous} have more similar neural responses during naturalistic narrative experience. These findings underscore the importance of understanding differences in neural representations underlying differences in perception and behavior across people. However, the full extent of variability within rich neural structure that gives rise to unique visual experience remains largely unknown.

Here, we investigated the dimensionality of neural representational structure underlying individual differences. Specifically, we quantified the extent to which inter-individual variability can be characterized through latent codes spanning multiple dimensional ranges. In neural population coding, dimensionality refers to the minimum number of axes needed to specify each response's position in representational space \cite{FUSI201666}. A prevailing view in cognitive neuroscience holds that cortical representations embed complex, high-dimensional stimuli into lower-dimensional latent subspaces to support generalization \cite{Lehky_Kiani_Esteky_Tanaka_2014, dicarlo2007untangling}. However, recent studies leveraging large-scale neural datasets show that natural stimulus representations in visual cortex span at least thousands of dimensions, with dimensionality scaling unboundedly up to the limit of the data \cite{stringer2019high, gauthaman2024}. This observation suggests that neural representations are more high-dimensional than previously appreciated. Crucially, dimensions beyond the leading, high-variance components encode stimulus information that may be relevant for perception but is not well-characterized by conventional analyses \cite{gauthaman2024}. These findings raise a critical question: Are there reliable individual differences across many latent dimensions of neural responses during naturalistic visual experience, or are they concentrated in a subset of dimensions that explain most of the variance in neural data? If neural representations are organized across multiple scales of variance, then individual differences in these representations might similarly span many latent dimensions. Alternatively, although stimulus-related variance is embedded in high-dimensional subspaces, it is possible that only a low-dimensional subspace might be relevant for individual differences in visual experience \cite{shine2019human, misra2021learning, lee2023hyper}. 

In this work, we use cross-decomposition to examine individual differences across the full spectrum of dimensions underlying cortical population codes. Our method combines spectral analysis with cross-validated functional alignment to characterize how neural representations vary between individuals at multiple representational scales. Using fMRI data of subjects viewing naturalistic movies, we quantify both the reliability and behavioral significance of intersubject variability in movie-evoked neural responses throughout the visual cortical hierarchy. Our findings reveal that reliable individual differences in visual processing manifest as distinct coactivation patterns distributed across multiple scales of neural representation, from dominant patterns explaining large amounts of variance to subtler patterns explaining increasingly smaller amounts. Crucially, we find that intersubject variability patterns are qualitatively distinct at each dimensional range and cannot be explained by conventional intersubject correlation (ISC) measures \cite{hasson2004intersubject}, revealing unique aspects of individual neural processing distributed across multiple levels of granularity. These multidimensional neural response patterns predict individual differences in movie recall, with neural-behavioral correlations distributed over a wide range of representational dimensions. These results highlight the multiscale structure of individual differences and establish a new spectral framework that advances our understanding of the high-dimensional neural architecture underlying subjective visual experience.

\section{Results}

\subsection{High-dimensional structure in cortical responses to movies}
One key to understanding neural population codes is characterizing how variance is distributed across latent dimensions of neural activity. Traditional approaches like principal component analysis (PCA) decompose neural responses into orthogonal dimensions sorted by variance. However, PCA captures both stimulus-driven and noise-related variance, potentially obscuring reliable stimulus-evoked patterns. Here, we employ cross-decomposition \cite{gauthaman2024} to isolate reliable stimulus-related variance in neural activity. Our approach isolates stimulus-dependent activations through cross-validation: basis sets derived from training data are used to compute covariance spectra on held-out test data. Like hyperalignment \cite{haxby2011common, guntupalli2016model}, cross-decomposition performs the Procrustes transformation to map a pair of neural response data into a set of orthogonal latent dimensions that maximize the covariance between the data sets. \added{We then obtain cross-validated covariance estimates for these aligned dimensions.} Importantly, \added{our method characterizes how information is distributed across the full spectrum of dimensions, allowing us to reveal the contribution of many subtle higher-rank dimensions that are typically overlooked.}
% \added{The method then characterizes the dimensionality of this shared space by quantifying cross-validated covariance across the full spectrum of aligned dimensions.} Importantly, \added{while hyperalignment focuses on cross-subject alignment achieved by finding shared representational spaces, }\comment{Reviewer \#2, Q4; Reviewer \#3, Q1}our method provides a spectral characterization of shared variance, decomposing the shared signal into latent dimensions ordered by explained covariance.

We hypothesized that stimulus-related variance in movie-evoked neural responses would be distributed across orders of magnitude of latent dimensions. While previous work used cross-decomposition to estimate the dimensionality of neural representations evoked by natural images \cite{gauthaman2024}, it has not been previously applied to dynamic, naturalistic movie responses. Here we applied cross-decomposition to a dataset of \replaced{42}{43} subjects viewing four naturalistic movies in an fMRI scanner \cite{savasegal}. The dataset was originally curated to study individual differences in neural event segmentation during movie viewing and contained movie stimuli with highly varied content. Thus, these movies provide an excellent stimulus set for testing the generalization of our findings using held-out stimuli that are qualitatively distinct from the training stimuli. Specifically, we used a one-movie-out cross-validation scheme, which provided a stringent test of reliability. 

\begin{figure}
	\centering
        \includegraphics[width=1\textwidth]{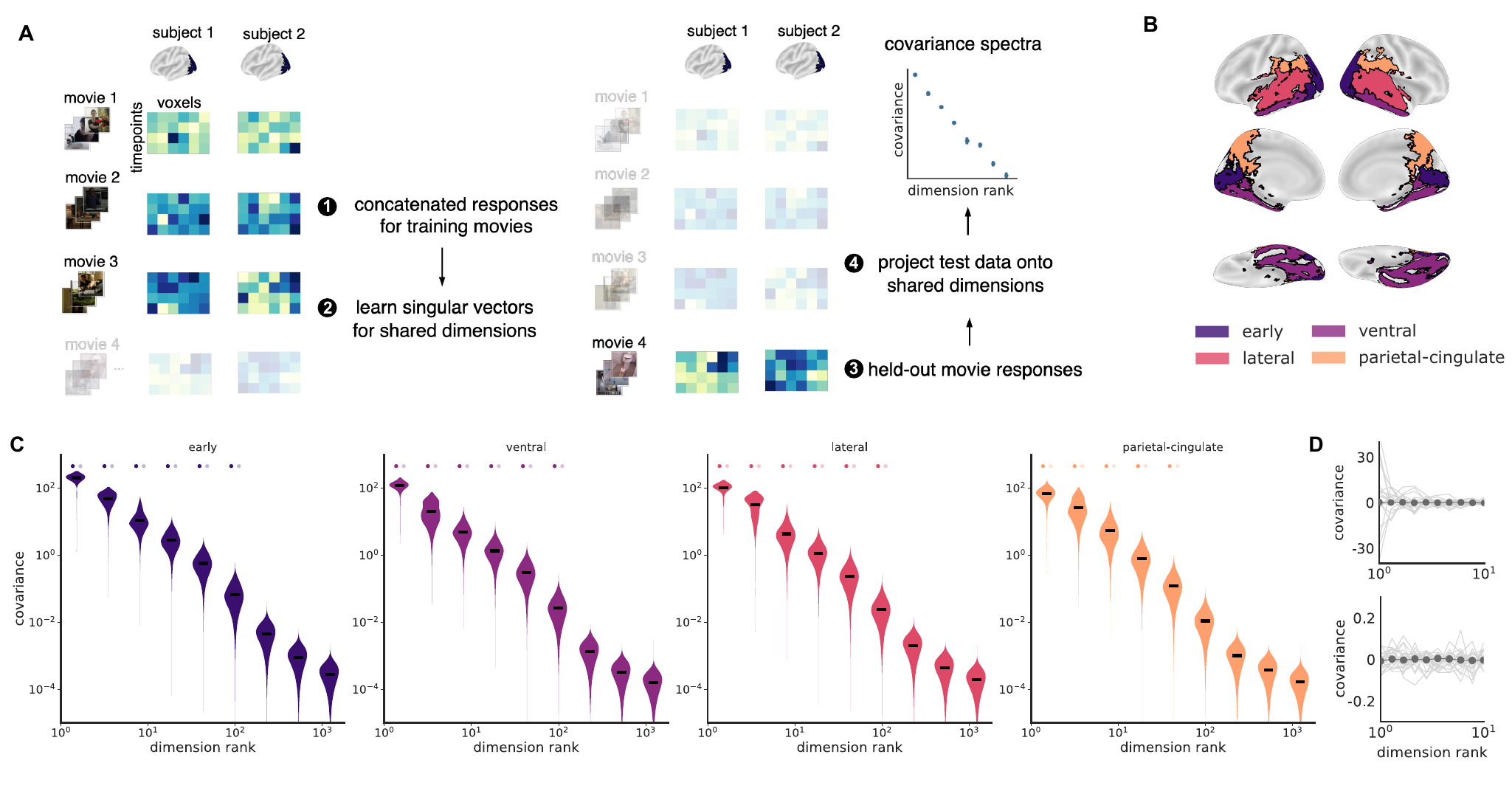}

	\caption{\textbf{Cross-decomposition analysis reveals many shared neural dimensions during movie viewing.}
		 (\textbf{A}) One-movie-out cross-validated cross-decomposition is used to identify reliable neural signal. Left: Shared dimensions from a subject pair are extracted from neural responses to three movies, with one movie held-out for testing. Right: The held-out movie responses are mapped onto these shared dimensions ranked by the covariance on the training set, generating a covariance spectrum that quantifies the amount of reliable signal on the test data. The procedure is repeated for each held-out movie to yield an average spectrum for each subject pair. (\textbf{B}) Regions of interest (ROIs) spanning the cortical hierarchy from low-level vision to high-level semantics: early visual areas (purple), ventral temporal (violet), lateral temporal (red), and posterior parietal-cingulate (orange) regions. (\textbf{C}) Cross-validated covariance spectra for each ROI. Violin plots show the distribution of reliable covariance across all subject pairs. The reliable stimulus-related covariance follows a \replaced{heavy-tailed}{power-law} distribution across dimension-rank in all regions, demonstrating that individual neural response patterns span multiple orders of magnitude rather than being confined to low-dimensional subspaces. \added{Dots above each violin indicate statistical significance ($p < 0.05$, Bonferroni-corrected). Significance was tested against two baseline distributions: darker dots indicate significance against the randomly block-shuffled baseline, and lighter dots indicate significance against the random rotation baseline.} (\textbf{D}) \added{Example null spectra for early visual cortex. These plots show sample traces of individual null spectra (grey traces) and the mean of null spectra (dark line). These plots illustrate that individual null spectra fluctuate around zero, and they show that in expectation, the null distributions are flat. Top: mean permutation null spectra are centered at zero, indicating that stimulus desynchronization eliminates the heavy-tailed covariance structure. Bottom: mean random rotation spectra are similarly flat and centered at zero, confirming that arbitrary rotation does not lead to the observed covariance pattern.}
         }
	\label{fig:1}
\end{figure}

\added{Cross-decomposition requires two independent neural response datasets to identical stimuli to isolate stimulus-driven variance from noise. Although the movies were not shown repeatedly, all subjects viewed the same movies, allowing us to perform cross-decomposition using the responses from pairs of subjects.} For a pair of subjects, we extracted shared latent dimensions from a training set of movies, with one movie held out for testing (Fig. \ref{fig:1}\textbf{A}, left). During the testing phase, the held-out movie responses from both subjects were projected onto shared latent dimensions, yielding a cross-covariance spectrum for the held-out data (Fig. \ref{fig:1}\textbf{A}, right). We averaged the cross-covariance spectra across movie-folds and repeated this procedure for every subject pair.

We applied this procedure to neural responses from a series of anatomically defined regions of interest (ROIs) spanning a processing hierarchy from low-level vision to high-level semantics, which we expect to be strongly modulated by movie stimuli (Fig. \ref{fig:1}\textbf{B}). These regions included early visual regions, ventral visual stream, lateral visual stream, and higher-level regions in the posterior parietal and cingulate cortex (See Methods). These ROIs allowed us to examine on how movie-evoked responses are represented across the cortical hierarchy, from early visual processing to multimodal regions associated with complex semantic integration. 

Movie-evoked responses showed a \replaced{heavy-tailed}{characteristic power-law} distribution of variance across latent dimensions in all examined regions, with cross-validated covariance spanning orders of magnitude of dimension ranks (Fig. \ref{fig:1}\textbf{C}). \added{Dimension ranks indicate the ordering of neural patterns from those explaining the most covariance (rank 1) to those explaining the least.} \deleted{The smooth decay in covariance values across ranks showed that movie-evoked information is distributed across many dimensions.} To verify that the covariance estimates over a wide range of ranks are statistically significant, we performed \replaced{two distinct null tests. Permutation testing disrupted temporal synchronization between subjects by randomly shuffling timepoint blocks, while random rotation testing applied arbitrary orthogonal transformations}{permutation tests by randomly block-shuffling movie-evoked responses across timepoints (See Methods)}. We confirmed that up to approximately 150 dimension ranks, the covariance statistics are significantly above \replaced{both null distributions in almost all ROIs ($p < 0.05$, Bonferroni-corrected; Fig. \ref{fig:1}\textbf{C}), demonstrating that the observed heavy-tailed decay is not expected by chance in either null distribution and reflects genuine stimulus-driven functional alignment.}{permuted spectra in all ROIs. We thus focus on this range of dimensions for our analyses} We also examined the characteristic spatiotemporal scales of covariance patterns captured by each dimension and found a trend going from coarse- to fine-scale granularity across dimension ranks in both space and time (Fig. \ref{figS2}). Thus, the cross-decomposition analysis showed that the naturalistic movie viewing responses contained many dimensions of meaningful variance shared between individuals beyond the first few components, with successive dimensions reflecting finer patterns of cortical activity.

\subsection{Individual differences in high-dimensional cortical response patterns}
Having established that there is reliable variance spanning many dimensions, we next investigated whether these high-dimensional response patterns contain reliable information about individual variability. Although high-rank dimensions contain reliable stimulus-related signal, most variance in the data is concentrated in the first few dimensions. Thus, high-rank dimensions might not capture meaningful ways in which individuals differ in their neural processing. Alternatively, despite their decreased variance, higher-rank dimensions might nonetheless contain unique information about individual differences in neural processing that cannot be captured by the dominant lower-rank dimensions. 
 
\begin{figure} 
	\centering
	\includegraphics[width=1\textwidth]{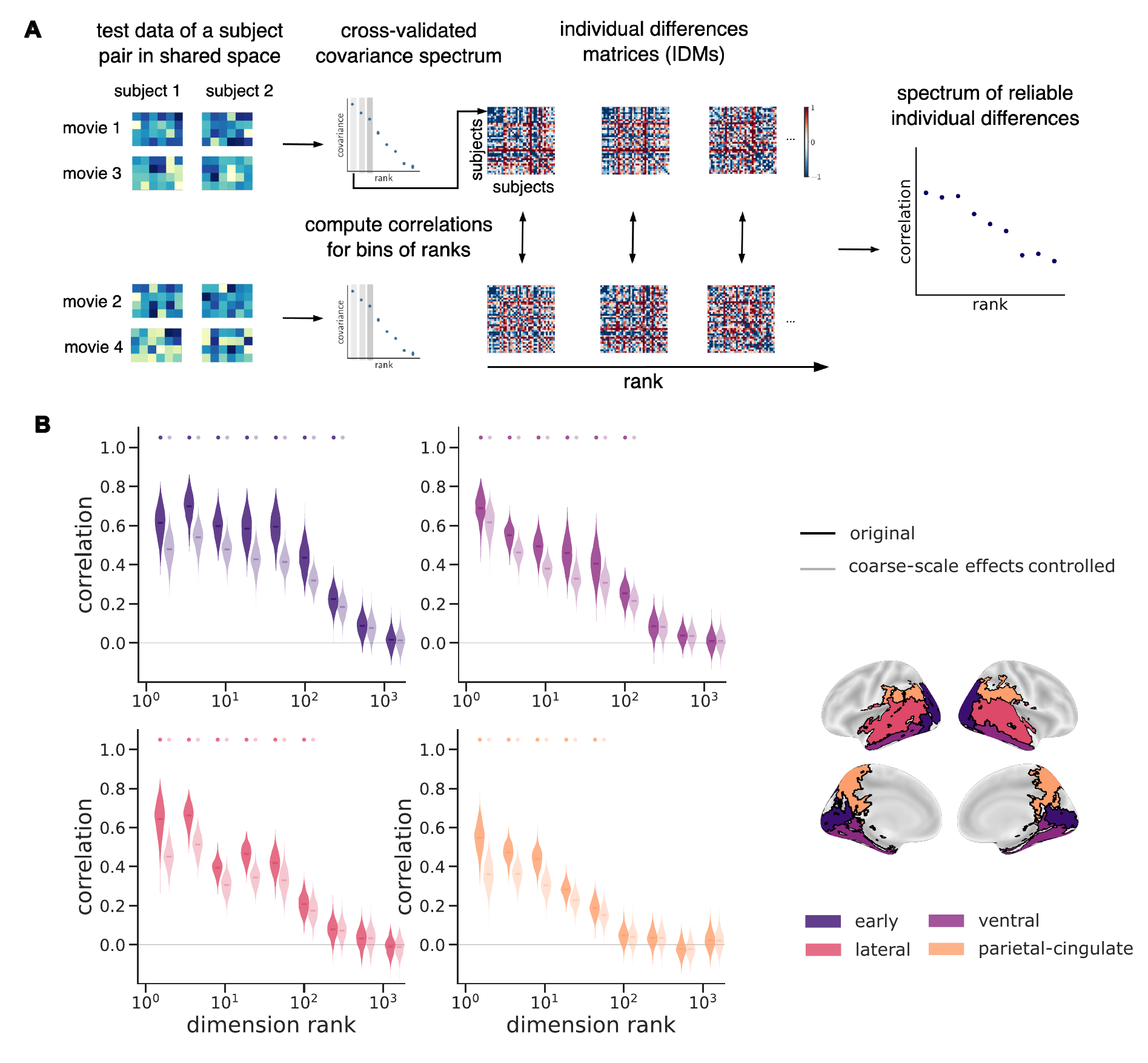}
	\caption{\textbf{Reliable individual differences persist across multiple dimensions of shared neural representations during movie viewing.}
		 (\textbf{a}) Individual differences analysis compares patterns between independent sets of data to assess reliability. Covariance spectra are generated separately for neural responses to odd and even movies from each subject pair. These spectra are used to construct rank-wise individual differences matrices (IDMs), where red/blue colors indicate high/low dissimilarity between subjects (covariance values normalized for visualization purposes). Correlations between odd and even IDMs quantify reliability of individual difference patterns at each dimension rank, as shown in the hypothetical correlation plot on the right. (\textbf{b}) The observed correlations in all ROIs show significance across \deleted{all examined }dimension ranks up to two orders of magnitude. The persistence of significant correlations across high-dimensional ranks indicates that fine-grained neural response patterns in cross-validated covariance spectra contain reliable individual differences that cannot be reduced to simpler, coarse-scale properties detected by voxelwise intersubject correlation (ISC) measures. Violin plots show the bootstrap distribution of correlations between odd and even movie IDMs for each dimension rank bin, with 95\% bootstrap confidence intervals (n=1,000 resamples). Dark shades represent correlations between original IDMs; lighter shades show correlations after controlling for coarse-scale effects through partialing out voxelwise ISC IDMs. \added{Grey horizontal line indicates zero correlation.} Dots above each violin indicate statistical significance ($p < 0.05$, Bonferroni-corrected).}
	\label{fig:2}
\end{figure}

To investigate this, we characterized patterns of individual differences and assessed their reliability across different ranges of dimensions (Fig. \ref{fig:2}\textbf{A}). We captured subject-by-subject similarity patterns of shared information in neural responses using individual differences matrices (IDMs) \cite{feilong2018reliable}. We constructed IDMs for bins of latent dimensions for separate subsets of movies and computed the split-half reliability of individual differences across the full dimensional spectrum. If reliable individual differences span the full spectrum, this would indicate that the entire range of dimensions in these high-dimensional codes are potentially relevant to how individuals represent naturalistic stimuli. The reliability of individual differences in neural representations is important because it establishes the upper bound of the correlation between individual differences in neural codes and individual differences in behavior \cite{spearman1961proof}.

We found that individual differences were reliable across different subsets of movie stimuli \replaced{up to a hundred dimensions in all ROIs (Fig. \ref{fig:2}\textbf{B}).}{throughout the entire spectrum of dimensions tested.} The reliability of individual differences, quantified as correlations between IDMs derived from different sets of movies, remained significantly above zero \replaced{across orders of magnitude of dimension ranks}{at all dimensional ranges} (bootstrapped confidence intervals, n=1,000 resamples; $p < 0.05$, Bonferroni-corrected). This reliability was highest in low-rank dimensions but persisted \replaced{across a wide range of dimensions in all ROIs}{even in high-rank dimensions}. \added{We additionally performed a whole-brain analysis that confirmed the strongest high-dimensional IDM correlations are in the posterior ROIs analyzed here (Fig. \ref{figS5}).}

These surprisingly stable individual differences in high-dimensional codes are brought out by the cross-decomposition procedure, which increases the detectability of stimulus-related signal and reveals fine-grained idiosyncrasies in neural responses. Specifically, functional alignment transforms individual subjects' data into a shared representational space where fine-grained, multivariate patterns of activity captured in latent dimensions can be directly compared across subjects \cite{haxby2011common}. To rule out the possibility that these high-dimensional effects can be explained by coarse response properties, we computed individual differences based on average voxelwise similarity and partialed out these similarity patterns from our IDMs at each dimensional range (see Methods). This control analysis isolates individual differences encoded in the multivariate pattern structure beyond what can be explained by coarse-grained voxel correlations.

While controlling for these coarse-scale effects modulated the magnitude of the IDM correlations, the overall pattern of reliable differences remained robust across the spectrum, and we continued to observe statistically significant reliability extending over a hundred dimensions in all regions (Fig. \ref{fig:2}\textbf{B}). \added{We performed an additional validation that successively partials out correlations from preceding dimensions and confirmed that high-rank dimensions contain reliable information not captured by lower ranks (Fig. \ref{figS3}).}
Thus, high-dimensional individual differences cannot be reduced to simpler, coarse-scale response properties but instead reflect differences in the fine-grained structure of neural representations. 

\begin{figure} 
	\centering	\includegraphics[width=1\textwidth]{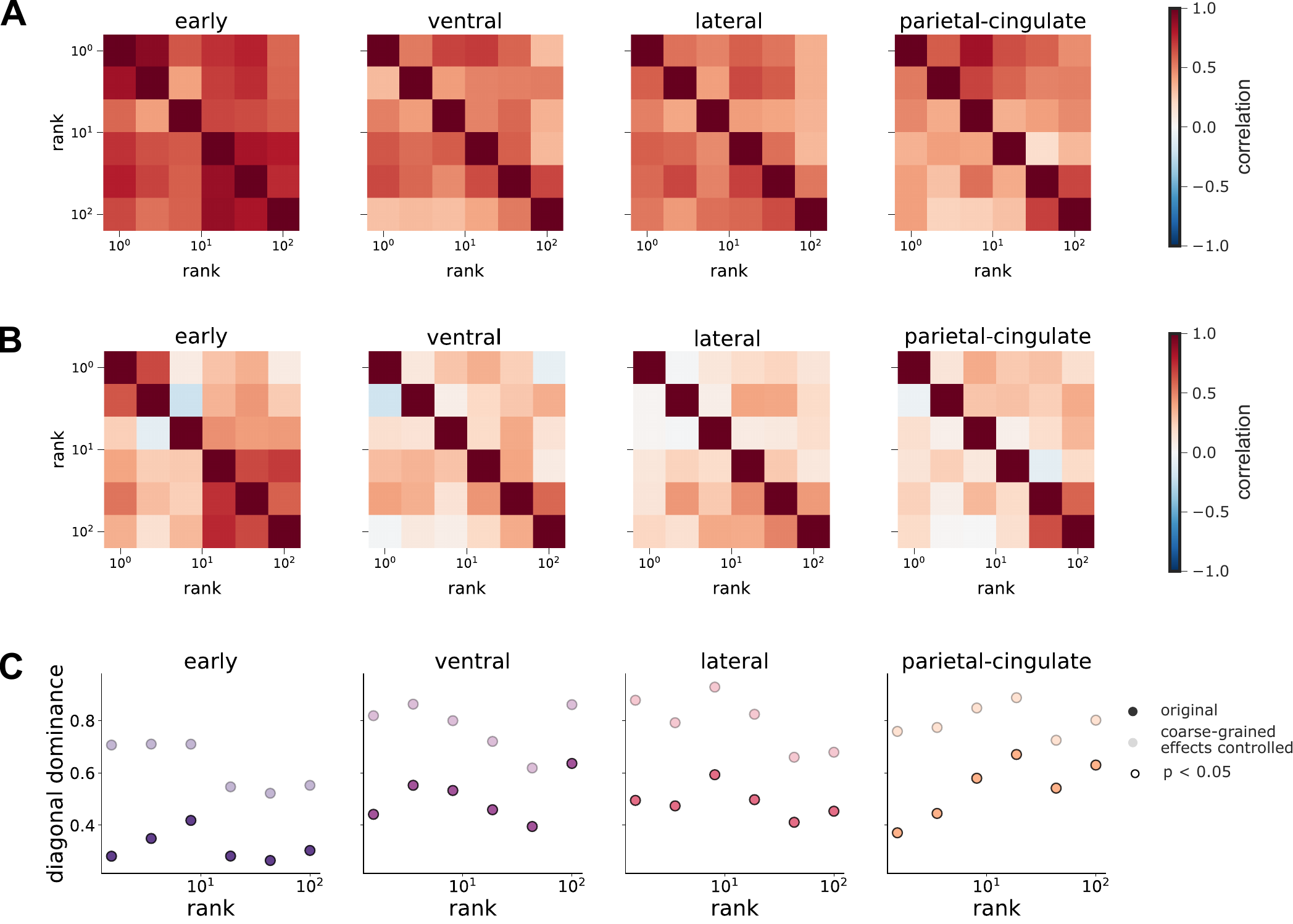}
	\caption{\textbf{Distinct patterns of individual differences emerge across different dimensional ranges.}
        (\textbf{a}) Correlation between individual difference matrices (IDMs) reveal stronger relationships between matching dimensional ranges than between different ranges. Each element represents the correlation between even-movie IDMs from one range (rows) and odd-movie IDMs from another range (columns)\added{, normalized by the reliabilities of the corresponding ranges such that diagonal values equal 1}. 
        \replaced{The prominent diagonal structure indicates that same-range correlations are consistently higher than different-range correlations}{Diagonal values show same-range correlations that are consistently higher than off-diagonal values}, demonstrating dimension-specific individual difference patterns. From left to right: early visual, ventral temporal, lateral visual, and posterior parietal-cingulate regions.
        (\textbf{b}) Same analysis after controlling for coarse-scale effects by partialing out voxelwise intersubject correlation (ISC) IDMs. The persistent diagonal structure indicates distinct, dimension-specific information in the residual, fine-grained patterns of individual differences. 
        (\textbf{c}) Diagonal dominance (difference between on-diagonal correlation and mean off-diagonal correlation) remains significant across dimensional ranges before and after removing the coarse-grained effects. The diagonal dominance across ranks demonstrates that each dimensional range captures distinct aspects of individual neural processing during naturalistic viewing. Solid points show values from original matrices in \textbf{a}; lighter points show values after controlling for coarse-scale effects detected by conventional measures. All points shown are statistically significant (permutation tests, $p < 0.05$). }
	\label{fig:3}
\end{figure}

After establishing that individual differences are reliable across many ranks, we next asked: do different dimensional ranges capture similar or distinct patterns of individual differences? If each range of dimensions reflects unique aspects of individual neural processing, we would expect the correlation between IDMs from the same dimensional range to be significantly stronger than correlations between different ranges. To test this hypothesis, we computed correlations between IDMs derived from different dimensional ranges across different sets of movies. \added{We normalized each correlation by the geometric mean of the within-range reliabilities. Since dimensional bins above rank 100 did not show significant reliability, we excluded them from these analyses.} The resulting correlation matrices revealed a prominent diagonal structure across all cortical regions. The strongest correlations were observed between matching dimensional ranges, with the mean correlation along the diagonal significantly higher than off-diagonal elements (Fig. \ref{fig:3}\textbf{A}). This pattern was particularly pronounced after controlling for coarse-grained response properties (Fig. \ref{fig:3}\textbf{C}), revealing fine-grained patterns of variability that were obscured by coarse-scale effects. Furthermore, region-to-region comparisons showed that removing the coarse-grained patterns of intersubject correlations also brings out ROI-specific patterns of variability at each dimensional range (Fig. \ref{figS4}). These results indicate that each range of latent dimensions not only contains reliable individual differences but captures qualitatively distinct patterns of individual neural processing. This dimensional specificity reveals that individuals meaningfully differ on multiple scales of naturalistic stimulus processing.

\subsection{Individual differences in cortical population codes and the interpretation of movies}
Our analysis demonstrated that reliable individual differences span over a hundred dimensions in both low-level and high-level cortical regions. We next sought to understand whether these high-dimensional neural response patterns are behaviorally relevant to how individuals interpret movies. The dataset examined here is particularly well-suited for addressing this question because it includes behavioral data specifically designed to capture diverse interpretations of naturalistic stimuli. 

\begin{figure} 
	\centering
	\includegraphics[width=1\textwidth]{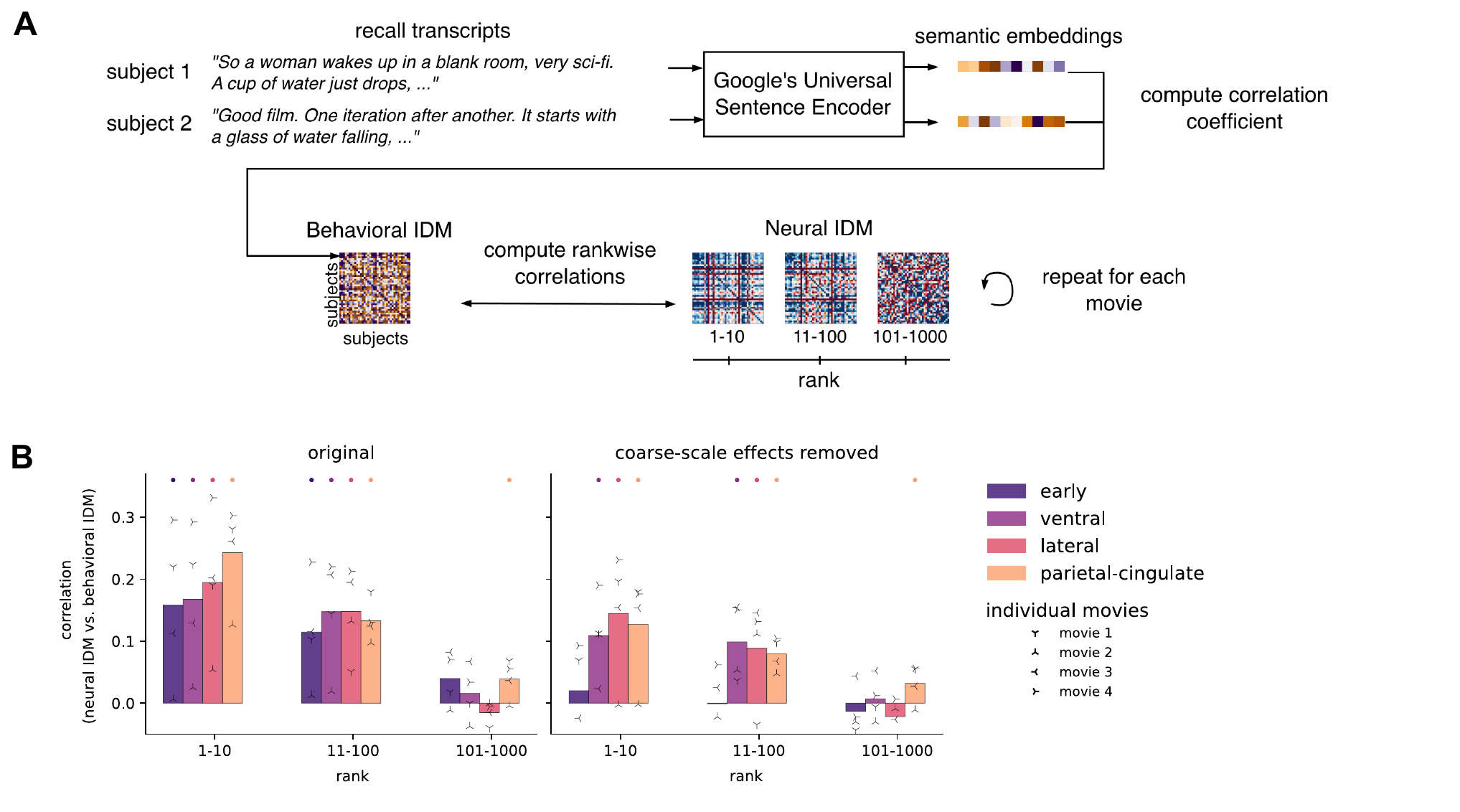}
	\caption{\textbf{High-dimensional neural patterns predict individual differences in subjective movie interpretations.}
		 (\textbf{A}) Neural-behavioral correlation analysis links brain activity patterns to subjective experiences. Free recall transcripts from each subject are transformed into semantic embeddings using Google's Universal Sentence Encoder, generating subject-by-subject behavioral individual difference matrices (behavioral IDMs). These matrices capture similarities in how individuals interpreted and described the movies. Behavioral IDMs are then correlated with neural IDMs computed at different dimensional scales \replaced{(1-10, 11-100, 101-1000)}{(1-10 and 11-100)}. Values of behavioral and neural IDMs are normalized for visualization purposes; Yellow/purple and red/blue colors indicate high/low dissimilarity between subjects in behavioral and neural IDMs, respectively. (\textbf{B}) Correlations between behavioral and neural patterns across cortical regions from early visual to posterior parietal-cingulate areas remain significant throughout multiple ranges of dimensions. Left panel shows raw correlations; right panel shows correlations after controlling for coarse-scale effects by partialing out voxelwise inter-subject correlations. Higher-level regions (ventral, lateral, parietal-cingulate) maintain significant correlations at higher dimensional ranges even after controlling for coarse-scale effects. Dots above each bar indicate statistical significance ($p<0.05$, FDR-corrected). Small triradiate symbols represent results for individual movies without averaging.}
	\label{fig:4} 
\end{figure}

To assess behavioral interpretations, we analyzed free recall transcripts collected after movie viewing. We used Google's Universal Sentence Encoder (USE) to transform these transcripts into semantic embeddings that capture the conceptual content of each subject's recall (Fig. \ref{fig:4}\textbf{A}). For each movie, we constructed behavioral IDMs by computing pairwise correlations between subjects' semantic embeddings, yielding a representation of the similarity structure in how individuals described and interpreted what they had seen. \added{Notably, the average between-subject correlation for the same movie was 0.59 ± 0.10, compared to the average within-subject correlation across different movies of 0.47 ± 0.09. This shows that the divergence across subjects for the same movie is nearly as strong as the divergence across different movies within each subject, confirming substantial inter-individual variability in movie interpretation.} To test whether neural response patterns across different dimensional ranges predict behavioral similarities, we computed correlations between neural IDMs for each decade of latent dimensions and the behavioral IDMs. \added{Because neural-behavioral correlations are more subtle than IDM reliability correlations, we used coarser dimensional binning to improve statistical power while still capturing distinctions across the dimensional spectrum.} 

We found that neural similarity patterns predict behavioral similarity across dimensional ranges spanning two orders of magnitude in all regions examined (Fig. \ref{fig:4}\textbf{B}, left panel). These brain-behavior correlations remained significant up to dimensions in the 11-100 range across all cortical regions ($p < 0.05$, FDR-corrected). When controlling for coarse-scale effects, the correlation between neural and behavioral IDMs was attenuated across all regions but remained significant in the 11-100 dimensional range for the ventral, lateral, and posterior parietal-cingulate regions (Fig. \ref{fig:4}\textbf{B}, right panel). In early visual cortex, the correlation between neural and behavioral IDMs was substantially reduced and no longer significant. Thus, while occipital regions contain coarse-grained information related to behavioral similarity, they lack fine-grained patterns that meaningfully correlate with individual differences in movie recall. In contrast, higher-level regions maintained behaviorally relevant information in fine-grained activity patterns after controlling for global response properties, suggesting that these subtle neural variations are important for capturing individual differences in stimulus interpretation. \added{Partial correlation analyses controlling for other dimensional ranges further confirmed that different ranges of dimensions provided independent contributions to this relationship (Fig. \ref{figS6}\textbf{A}).}

Together, the persistence of brain-behavior correlations across multiple dimensional ranges demonstrates that behaviorally relevant individual differences extend well beyond the dominant dimensions that explain the majority of variance in neural responses. These findings reveal that each subject's unique perspective on naturalistic content is reflected in high-dimensional population codes distributed over a hundred dimensions in each region of the cortical hierarchy, suggesting that fine-grained neural response patterns typically overlooked in conventional analyses may be crucial for capturing the rich diversity of individual visual experiences.

\subsection{Neural alignment across high-dimensional codes predicts concrete narrative descriptions}

\begin{figure} 
	\centering
	\includegraphics[width=1\textwidth]{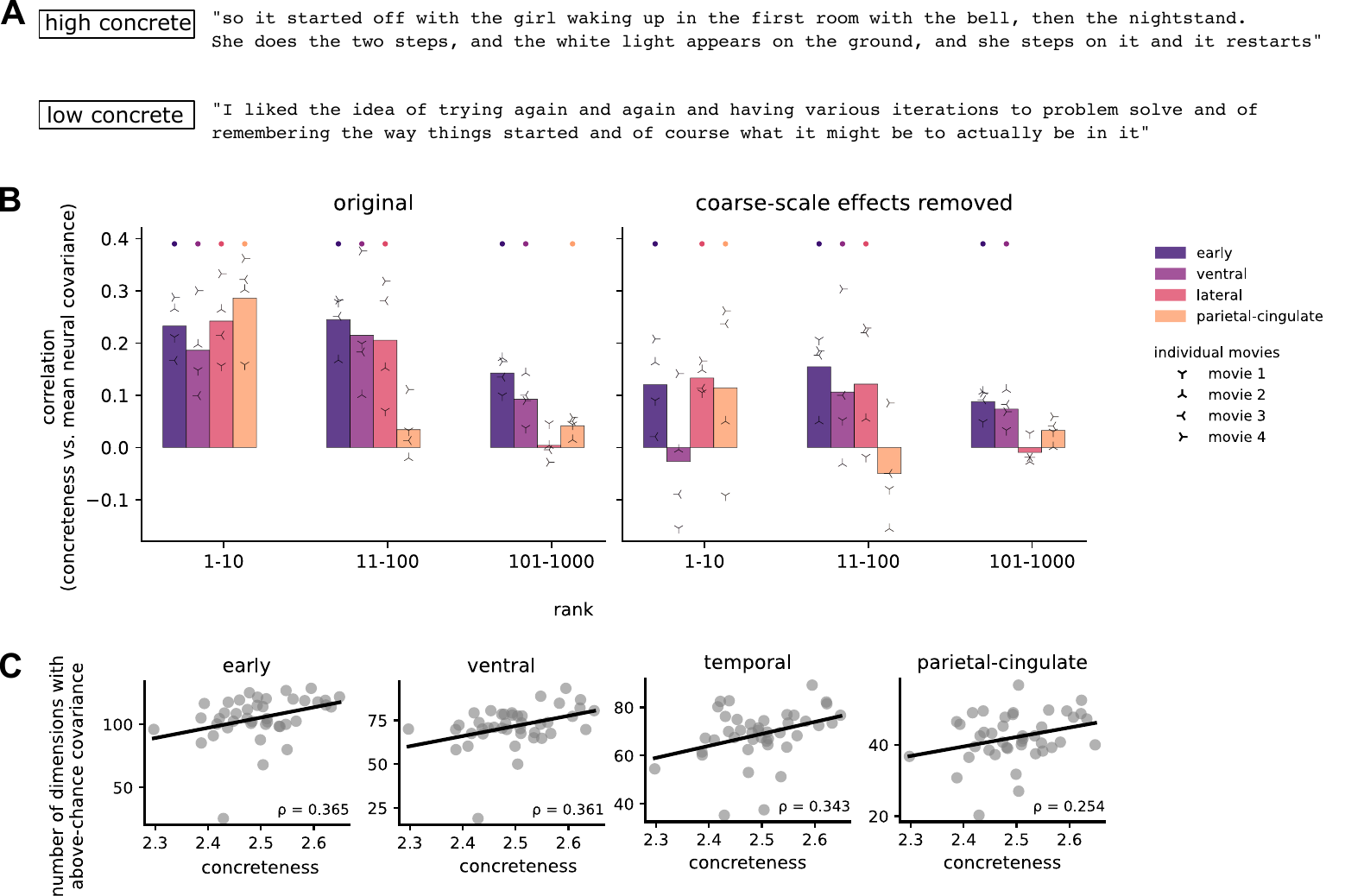}
	\caption{\added{\textbf{Concrete narrative descriptions predict greater neural alignment across high-dimensional codes.}
		 (\textbf{A}) Sample excerpts from two subjects illustrating the range of narrative concreteness, from highly concrete descriptions emphasizing perceptible events and objects (top) to more abstract, thematic accounts (bottom). (\textbf{B}) 
         Individuals who produced more concrete narrative descriptions showed greater neural similarity to others across multiple dimensional ranges. Bar plots show correlations between individual concreteness scores and mean intersubject neural covariance across dimensional ranges. Left: original covariance; right: after removing coarse-scale effects. Dots above each bar indicate statistical significance ($p<0.05$, FDR-corrected). Small triradiate symbols represent results for individual movies without averaging. (\textbf{C}) 
         Subject-level correlations between narrative concreteness and the number of shared neural dimensions with above-chance inter-subject covariance. Individuals with more concrete narrative styles tended to have higher shared neural dimensionality across all cortical regions (Spearman $\rho=0.254-0.365$). Each point represents one subject.}}
	\label{fig:5} 
\end{figure}

\added{Having established that high-dimensional neural patterns predict individual differences in movie recall, we next sought to understand what interpretable aspects of behavior might relate to these individual differences in neural representations. Inspection of subjects' verbal recall transcripts revealed an intriguing pattern in descriptive style. While some subjects primarily focused on concrete, perceptual details of the movies such as specific actions, objects, and events, others reported more abstract, interpretive aspects such as themes, valence, and personal reactions (see examples in Fig. \ref{fig:5}\textbf{A}). This observation led us to hypothesize that the degree of concreteness in subjects' descriptions might be systematically related to their neural similarity.}

\added{We reasoned that individuals who encode and recall concrete details would show greater neural alignment with one another because concrete details reflect more consistent, stimulus-driven processing of the shared sensory input. In contrast, abstract and interpretive processing may be more idiosyncratic, leading to greater inter-individual variability in neural responses. To test this hypothesis, we quantified the concreteness of each subject's recall transcripts using a large-scale concreteness ratings database \cite{brysbaert2014concreteness} and correlated it with their average neural covariance with all other subjects across different dimensional ranges (see Methods). We found that individuals who produced more concrete narrative descriptions showed significantly greater neural similarity to others across multiple dimensional ranges (Fig. \ref{fig:5}\textbf{B}).}

\added{We then examined whether individuals who recall more concretely had higher shared dimensionality. We computed the number of significant shared dimensions between each subject and all others and correlated this with each subject's concreteness rating. Individuals with more concrete narrative styles tended to show higher shared neural dimensionality across all cortical regions examined (Fig. \ref{fig:5}\textbf{C}). These findings suggest that an important factor underlying individual differences in both neural representations and behavioral recall is the degree to which subjects encode the rich concrete details of their experiences versus more abstract, interpretive aspects.
}

\section{Discussion}
By applying cross-decomposition to fMRI data collected during movie viewing, we characterized the complex representational structure underlying individual differences. Strikingly, we found a remarkable persistence of reliable individual differences across this expansive dimensional spectrum. Our analyses revealed that even dimensions with relatively low variance contain stable information about individual neural processing. The dimensional specificity observed in patterns of individual differences provides compelling evidence for multiple orthogonal axes of variation in how individuals process identical visual stimuli. This dimensional specificity persisted even after controlling for coarse-scale response properties captured by voxelwise ISC measures, confirming that fine-grained multivariate patterns contribute substantially to neural individuality beyond what can be explained by coarse-grained responses. Lastly, the preservation of significant correlations between neural and behavioral similarity matrices across dimensional ranges further demonstrated that these high-rank dimensions, despite explaining relatively little overall variance, encode functionally meaningful information that shapes individual interpretations of visual input. \added{Notably, individuals who recall more concrete details exhibit higher-dimensional shared representations.}

These findings bridge recent computational advances with fundamental questions about neural coding. Large-scale modeling efforts \cite{krizhevsky2017imagenet, he2016deep, dosovitskiy2020image, vaswani2017attention, brown2020language, kaplan2020scaling} and big-data approaches in neuroscience \cite{allen2022massive, hebart2023things, manley2024simultaneous} have enabled empirical estimates of latent dimensionality that are orders of magnitudes greater than previously possible. Recent work in systems neuroscience found that cortical representations are high-dimensional \cite{stringer2019high, contier2024distributed, elmoznino2024high, manley2024simultaneous, gauthaman2024}, in line with theoretical perspectives suggesting that high-dimensional neural codes confer computational advantages for complex tasks \cite{FUSI201666, sorscher2022neural, elmoznino2024high}. Our work extends this framework to the neural basis of individual differences and\added{, contrasting with earlier findings of organization along a small number of dimensions based on leading high-variance components \cite{huth2012continuous, tarhan2020sociality, khosla2022highly},} demonstrates that neural population codes leverage their high-dimensional capacity to organize shared and idiosyncratic features across multiple representational ranges. Critically, our findings connecting high-dimensional neural patterns to individual differences in movie recall suggest that subjective experience emerges not from dominant neural signals alone, but from the integration of information distributed across the entire variance spectrum.

\added{Our findings that reliable neural representations span at least a hundred dimensions within individual regions are particularly striking given that this high-dimensional structure extends throughout the visual cortical hierarchy, including higher-level semantic regions in posterior parietal and cingulate cortex. The richness of naturalistic movie stimuli, combined with the cross-decomposition method, enables us to detect this expansive representational structure across both early sensory and higher-order integrative regions.} \added{These results are consistent with, and extend, prior functional alignment methods that have identified shared representations spanning hundreds of latent dimensions across the whole cortex.}\cite{guntupalli2016model} However, while previous work has used hyperalignment to improve the detection of fine-grained individual differences \cite{haxby2011common, guntupalli2016model, feilong2018reliable, Guntupalli_2020, Feilong_Guntupalli_Haxby_2021}, these previous studies did not examine the range of latent dimensions that contribute to these individual differences. Our spectral approach, which examines stimulus-related variance across the full spectrum of latent dimensions, reveals that individual differences manifest uniquely at each dimensional range—from dominant patterns in low-rank dimensions to subtle variations in high-rank dimensions. By explicitly modeling the distribution of variance across the full dimensional spectrum, our analysis reveals that high-rank dimensions encode qualitatively distinct and behaviorally significant information. Thus, our work suggests focusing on the distribution and organization of variance across dimensions and provides a framework for characterizing the multiscale structure of neural individuality during naturalistic perception.

\added{Our findings suggest several important directions for future work. First, we found that reliability of neural individual differences based on higher ranks are relatively lower than that based on lower ranks. While several observations indicate lower effective signal-to-noise ratio at higher ranks, it remains to be investigated whether the lower reliabilities reflect decreased signal quality or genuinely smaller individual differences at finer representational scales. Second, our temporal autocorrelation analysis suggests that higher-rank dimensions operate at finer temporal scales, raising the question of whether higher ranks might encode stimulus information that varies more rapidly. Third, an exciting direction for future work will be to investigate whether specific dimensions or ranges of dimensions can be mapped onto interpretable features of the movie stimuli, or whether they have complex mixed selectivities that cannot be easily interpreted. Lastly, our current approach uses anatomically-defined ROIs and extracts shared components separately for each subject pair. Future extensions could use functionally-defined ROIs based on individualized cortical topographies and template-based alignment methods that enable comparison of the same components across all subject pairs, as implemented in the Individualized Neural Tuning model \cite{feilong2023individualized}.}

Together, our findings reveal that individual differences in visual processing manifest through high-dimensional neural codes distributed across multiple representational scales throughout the cortical hierarchy. This multiscale organization calls for approaches that capture the full spectrum of neural variability to comprehensively characterize the neural basis of individualized perceptual experience. This geometric perspective reframes our understanding of neural individuality---not as variation along a few principal dimensions, but as a complex, high-dimensional pattern of idiosyncrasies distributed throughout representational space. Such a framework offers promising new avenues for investigating individual differences in perception, cognition, and potential alterations in clinical populations. 

\newpage

%%%%%%%%%%%%%%%% MATERIALS AND METHODS %%%%%%%%%%%%%%%
\section*{STAR Methods}
\subsection{Key resources table}

\begin{table}[h]
\centering
\begin{tabular}{llp{6cm}}
\hline
REAGENT or RESOURCE & SOURCE & IDENTIFIER \\
\hline
\multicolumn{3}{l}{\textbf{Deposited Data}} \\
Naturalistic movie fMRI dataset & Sava-Segal et al.\cite{savasegal} & OpenNeuro: ds004516 \\
\hline
\multicolumn{3}{l}{\textbf{Software and Algorithms}} \\
fMRIPrep & Esteban et al.\cite{Esteban2019} & RRID:SCR\_016216 \\
FSL & Smith et al. \cite{smith2004advances} & RRID:SCR\_002823 \\
Python & Python Software Foundation & \url{https://www.python.org} \\
Google Universal Sentence Encoder & Cer et al.\cite{cer2018universal} & \url{https://tfhub.dev/google/universal-sentence-encoder/4} \\
custom code & This paper & \url{https://github.com/kelseyhan-jhu/idiosyncratic-neural-geometry} \\
\hline
\multicolumn{3}{l}{\textbf{Other}} \\
AAL3 Atlas & Rolls et al. \cite{ROLLS2020116189} & \url{https://www.gin.cnrs.fr/en/tools/aal/} \\
\hline
\end{tabular}
\end{table}

\subsection{Experimental model and study participant details}
\subsubsection{fMRI participants}
We analyzed publicly available fMRI data from Sava-Segal et al.\cite{savasegal}. From 48 initially recruited participants (all native English speakers and right-handed; 27 female, median age 24.5 years, range 19-64 years), the original study retained 43 participants with complete fMRI data for all four movies. We further excluded 1 participant with missing behavioral recall data, resulting in a final sample of 42 participants (all native English speakers and right-handed, 24 female, median age 24.5 years, range 19-64 years). All participants provided informed written consent under NIH Institutional Review Board approval.

\subsection{Method details}
\subsubsection{Movie stimuli}
Participants viewed four naturalistic audiovisual movie clips (duration 7:27-12:27 min). Movies varied in content and cinematic style, including social and affective elements and screen cuts. Three out of four movies depicted scenarios with social interactions (versus one with purely mechanical actions), and two contained screen cuts (versus continuous camera panning). Detailed stimulus descriptions and links are available in Sava-Segal et al.\cite{savasegal}.
\paragraph{Behavioral recall data}
Following each movie, participants provided free verbal recall for three minutes describing characters, events, and opinions. 

\paragraph{Narrative concreteness} \added{To quantify how concretely versus abstractly subjects described their experiences, we computed concreteness scores for each subject's recall transcripts using a large-scale database \cite{brysbaert2014concreteness}. This database provides 5-point scale ratings (1=abstract, 5=concrete) for 37,058 English words collected via Amazon Mechanical Turk. For each subject's transcript, we computed the mean concreteness rating across all words present in the database (mean coverage: 90.9\% of words per transcript, range: 85.4\%-96.3\%). We then averaged concreteness scores across all four movies for each subject to obtain a subject-level concreteness measure.}

\subsubsection{fMRI experiment}
\paragraph{fMRI data acquisition}
Functional MRI data were collected at the National Institutes of Health using a 3T Siemens Prisma scanner with a 64-channel head coil. Movies were presented in pseudorandomized order (counterbalanced across participants) via rear-projection screen viewed through an angled mirror using PsychoPy \cite{peirce2019}. Functional images were acquired using T2*-weighted multiband, multi-echo echo-planar imaging with TR = 1,000 ms, echo times = [13.6, 31.86, 50.12 ms], flip angle = 60°, 3.0 mm isotropic voxels, 52 slices (whole-brain coverage), multiband acceleration factor = 4. Anatomical images were acquired using T1-weighted MPRAGE with TR = 2,530 ms, TE = 3.30 ms, flip angle = 7°, 1.0 mm isotropic resolution. See Sava-Segal et al.\cite{savasegal} for complete acquisition details.

\paragraph{fMRIPrep}
% Functional images (TR=1 s) were preprocessed using fMRIPrep \cite{Esteban2019}, including motion correction, co-registration, and normalization to MNI space at 2mm isotropic resolution. 

Results included in this manuscript come from preprocessing performed using \emph{fMRIPrep} 21.0.2 \cite{fmriprep1, fmriprep2}, which is based on \emph{Nipype} 1.6.1 \cite{nipype1, nipype2}.

The next two sections (anatomical data and functional data) of boilerplate text were automatically generated by fMRIPrep with the express intention that users should copy and paste this text into their manuscripts unchanged. It is released under the CC0 license.

\paragraph{Anatomical data}
A total of 1 T1-weighted (T1w) images were found within the input BIDS dataset. The T1-weighted (T1w) image was corrected for intensity non-uniformity (INU) with \texttt{N4BiasFieldCorrection} \cite{n4}, distributed with ANTs 2.3.3 \cite{ants}, and used as T1w-reference  hroughout the workflow. The T1w-reference was then skull-stripped with a \emph{Nipype} implementation of the
\texttt{antsBrainExtraction.sh} workflow (from ANTs), using OASIS30ANTs as target template. Brain tissue segmentation of cerebrospinal fluid (CSF), white-matter (WM) and gray-matter (GM) was performed on the brain-extracted T1w using \texttt{fast} (FSL 6.0.5.1:57b01774 \cite{fsl_fast}). Volume-based spatial normalization to one standard space (MNI152NLin2009cAsym) was performed through nonlinear registration with \texttt{antsRegistration} (ANTs 2.3.3), using brain-extracted versions of both T1w reference and the T1w template. The
following template was selected for spatial normalization: \emph{ICBM 152 Nonlinear Asymmetrical template version 2009c}
\cite{mni152nlin2009casym} (TemplateFlow ID:
MNI152NLin2009cAsym).

\paragraph{Functional data}
For each of the 1 BOLD runs found per subject (across all tasks and
sessions), the following preprocessing was performed. First, a reference volume and its skull-stripped version were generated from the shortest echo of the BOLD run using a custom methodology of \emph{fMRIPrep}. Head-motion parameters with respect to the BOLD reference (transformation matrices, and six corresponding rotation and translation parameters) are estimated before any spatiotemporal filtering using \texttt{mcflirt} (FSL 6.0.5.1:57b01774) \cite{mcflirt}. BOLD runs were slice-time corrected to 0.452s (0.5 of slice acquisition range 0s-0.905s) using \texttt{3dTshift} from AFNI \cite{afni}. The BOLD time-series (including slice-timing correction when applied) were resampled onto their original, native space by applying the transforms to correct for head-motion. These resampled BOLD time-series will be referred to as \emph{preprocessed BOLD in original space}, or just \emph{preprocessed BOLD}. A T2* map was estimated from the preprocessed EPI echoes, by voxel-wise fitting the maximal number of echoes with reliable signal in that voxel to a monoexponential signal decay model with nonlinear regression. The T2*/S0 estimates from a log-linear regression fit were used for initial values. The calculated T2* map was then used to optimally combine preprocessed BOLD across echoes following the method
described in Posse et al.\cite{posse_t2s}. The optimally combined time series was carried forward as the \emph{preprocessed BOLD}. The BOLD reference was then co-registered to the T1w reference using \texttt{mri\_coreg} (FreeSurfer) followed by \texttt{flirt} (FSL 6.0.5.1:57b01774)\cite{flirt}, with the boundary-based registration cost-function \cite{bbr}. Co-registration was configured with six degrees of freedom. First, a reference volume and its skull-stripped version were generated using a custom methodology of \emph{fMRIPrep}. Several confounding time-series were calculated based on the \emph{preprocessed BOLD}: framewise displacement (FD), DVARS and three region-wise global signals. FD was computed using two formulations following Power (absolute sum of relative motions\cite{power_fd_dvars} and Jenkinson (relative root mean square displacement between affines\cite{mcflirt}. FD and DVARS are calculated for each functional run, both using their implementations in \emph{Nipype} following the definitions by Power et al.\cite{power_fd_dvars}).
The three global signals are extracted within the CSF, the WM, and the whole-brain masks. Additionally, a set of physiological regressors were extracted to allow for component-based noise correction \emph{CompCor}\cite{compcor}. Principal components are estimated after high-pass filtering the \emph{preprocessed BOLD} time-series (using a discrete cosine filter with 128s cut-off) for the two \emph{CompCor} variants: temporal (tCompCor) and anatomical (aCompCor). tCompCor components are then calculated from the top 2\% variable voxels within the brain mask. For aCompCor, three probabilistic masks (CSF, WM and combined CSF+WM) are generated in anatomical space. The implementation differs from that of Behzadi et al.~in that instead of eroding the masks by 2 pixels on BOLD space, the aCompCor masks are subtracted a mask of pixels that likely contain a volume fraction of GM. This mask is obtained by thresholding the corresponding partial volume map at 0.05, and it ensures components are not extracted from voxels containing a minimal fraction of GM. Finally, these masks are resampled
into BOLD space and binarized by thresholding at 0.99 (as in the
original implementation). Components are also calculated separately
within the WM and CSF masks. For each CompCor decomposition, the
\emph{k} components with the largest singular values are retained, such that the retained components' time series are sufficient to explain 50 percent of variance across the nuisance mask (CSF, WM, combined, or temporal). The remaining components are dropped from consideration. The head-motion estimates calculated in the correction step were also placed within the corresponding confounds file. The confound time series derived from head motion estimates and global signals were expanded with the inclusion of temporal derivatives and quadratic terms for each \cite{confounds_satterthwaite_2013}. Frames that exceeded a threshold of 0.5 mm FD or 1.5 standardised DVARS were annotated as motion outliers. The BOLD time-series were resampled into standard space, generating a \emph{preprocessed BOLD run in MNI152NLin2009cAsym space}. First, a reference volume and its skull-stripped version were generated using a custom methodology of \emph{fMRIPrep}. All resamplings can be performed with \emph{a single interpolation step} by composing all the pertinent transformations (i.e.~head-motion transform matrices,
susceptibility distortion correction when available, and
co-registrations to anatomical and output spaces). Gridded (volumetric) resamplings were performed using \texttt{antsApplyTransforms} (ANTs), configured with Lanczos interpolation to minimize the smoothing effects
of other kernels \cite{lanczos}. Non-gridded (surface) resamplings were performed using \texttt{mri\_vol2surf} (FreeSurfer).

Many internal operations of \emph{fMRIPrep} use \emph{Nilearn} 0.8.1
\cite{nilearn}, mostly within the functional
processing workflow. For more details of the pipeline, see the section corresponding to workflows in \emph{fMRIPrep}'s documentation.

\paragraph{Post-processing}
Preprocessed data were trimmed to movie segments and high-pass filtered (64 s cutoff) using FSL \cite{smith2004advances}.

\paragraph{Region of interest (ROI) definition}\label{roi}
We defined four anatomical regions of interest (ROIs) spanning the visual cortical hierarchy using the Automatic Anatomical Labeling atlas (AAL3) \cite{ROLLS2020116189}: the early visual regions (including calcarine, superior, middle, and inferior occipital cortices), the ventral visual stream (including lingual, fusiform, inferior temporal, and parahippocampal cortices), and the lateral visual stream (including middle and superior temporal cortices). We also created an ROI by combining higher-level regions in the posterior part of the cortex (including supramarginal and angular gyri, precuneus, and posterior cingulate cortex) that we collectively refer to as posterior parietal-cingulate ("parietal-cingulate"). We grouped these high-level posterior regions to maintain comparable voxel counts across ROIs while combining regions that support similar semantic and multimodal integration functions. The covariance spectra patterns show similar high-dimensional structure when these posterior regions are analyzed separately. \added{Maintaining similar ROI sizes was important because the number of detectable dimensions scales with the number of voxels.} ROI masks were applied to each subject's preprocessed movie fMRI responses to extract the corresponding voxel timeseries matrices. All regions included both left and right hemispheres.

\added{For whole-brain analysis (Supplementary Fig. \ref{figS5}), we defined 170 bilateral ROIs corresponding to all AAL3 parcels (Rolls et al., 2020). We excluded 11 ROIs where more than 10\% of subject pairs had insufficient overlapping voxels and thus insufficient rank to compute cross-covariance. The excluded regions were primarily subcortical structures with limited coverage in the functional data (anterior cingulate, thalamus, reuniens thalamic nucleus, medial geniculate nucleus, ventral tegmental area, and median raphe nucleus).}

\subsection{Quantification and statistical analysis}
\subsubsection{Estimating latent dimension}
\paragraph{Cross-decomposition}
We used cross-decomposition to estimate the dimensionality of the latent neural subspace shared by pairs of individuals. For each subject pair, we analyzed their data matrices $X, Y \in \mathbb{R}^{t \times v}$ representing $t$ timepoints and $v$ common voxels from subjects $i$ and $j$, respectively. \added{Here we used common voxels because the data were in MNI space, but this method does not require the two data matrices to be in a common space. Voxel timecourses were z-scored for each subject and each movie before computing cross-subject covariance.} We computed the training set cross-covariance eigenvalues and eigenvectors using singular vector decomposition: 
\begin{equation}
    \text{cov} (X_{train}, Y_{train}) = U \Sigma V^T
\end{equation}

\added{This singular value decomposition performs the Procrustes transformation, identifying the optimal orthogonal rotations ($U$ for subject $i$, $V$ for subject $j$) that align the two subjects' neural response patterns into a shared representational space that maximizes cross-covariance. The matrices $U$ and $V$ define the functional alignment between subjects. Then we projectected held-out test data onto the aligned dimensions (defined by $U$ and $V$ from training).} The cross-covariance eigenspectrum for the testing set was then derived as:
\begin{equation}
    \Sigma_{test}^{(i, j)} = \text{cov}\left(X_{test} U, Y_{test} V\right)
\end{equation}

\added{The test-set covariance spectrum, $\Sigma_{test}$, quantifies reliable stimulus-driven variance across dimension ranks. This test-set spectrum estimates dimensionality by identifying which dimensions carry reliable stimulus-driven signal as opposed to noise.}

This procedure was applied using leave-one-movie-out cross-validation across all subject pairs. Spectra for all test movies and subject pairs were aggregated by computing the mean covariance within logarithmically-spaced bins of latent dimensions. For each bin, we calculated a representative center value $c$ as the geometric mean of its edges and assigned each dimension rank to its corresponding bin. The binning approach allowed characterizing distributed aspects of neural representation.

\paragraph{Permutation testing}
Statistical significance was assessed through block permutation tests to maintain the temporal structure in fMRI data. We permuted \replaced{11}{6}-second blocks of \added{test }movie timeseries \added{independently for both subjects} and \replaced{projected the permuted data onto singular vectors learned from training data}{recomputed the cross-decomposition procedure} to generate a null distribution \added{from temporally desynchronized responses}. \added{This null distribution preserves the temporal autocorrelation structure of the data and thus provides a stringent test for whether observed covariance reflects meaningful stimulus-locked neural alignment. For each subject pair, we computed covariance spectra averaged across test movies and binned using the same logarithmic binning as observed data (n=1,000). At each dimensional bin, significance was assessed by comparing observed covariance against the permutation distribution ($p < 0.05$, Bonferroni-corrected). For Fig. \ref{fig:1}\textbf{D}, we show 10 individual permutation iteration traces from a randomly chosen single subject pair, along with the mean across all 1,000 permuted spectra for all subject pairs to illustrate the average null distribution centered at zero.} 

\paragraph{Random rotation testing}
\added{We performed additional random rotation control analyses to test whether any arbitrary rotation could produce the observed covariance pattern, as opposed to the specific Procrustes alignment that maximizes shared covariance. For each subject pair and test movie, we generated random orthogonal projections of the neural data and computed null covariance spectra. We generated random orthogonal projection matrices $Q_x \in \mathbb{R}^{v \times k}$ and $Q_y \in \mathbb{R}^{v \times k}$ for subjects $x$ and $y$, where $v$ is the number of voxels and $k$ is the number of components (set to maximum observed rank). These projection matrices were generated using QR decomposition of random Gaussian matrices. We projected each subject's standardized data into these random $k$-dimensional subspaces ($X_{rot} = X_{test} Q_x$, $Y_{rot} = Y_{test}Q_y$) and computed covariance spectra from the randomly projected data. We found that the rotation null distributions converged quickly, and because these analyses were computationally intensive, we decided to stop the procedure after 100 iterations. On each iteration, the covariance statistics were movie-averaged, binned, and aggregated across subject pairs as with the observed data. Significance was assessed by comparing observed covariance against the random rotation distribution ($p < 0.05$, Bonferroni-corrected).}

\paragraph{Spatial visualization}
We visualized the spatial loadings of latent components on cortical surface models to examine how neural patterns change across dimension ranks. For each region, we extracted the left singular vectors ($U$) of a sample pair from the cross-decomposition analysis, which represent the spatial loadings of each dimension. These loading patterns from individual dimensions were visualized with consistent thresholding (0.0001) and spatial smoothing (FWHM=2mm). Region-specific views were used to optimally display each ROI: medial views for posterior parietal-cingulate regions, ventral views for ventral temporal regions, and lateral views for early visual and lateral temporal regions.

\paragraph{Temporal autocorrelation analysis}
To quantify the temporal dynamics of different dimensional ranges, we computed autocorrelation functions for the time series associated with each latent dimension. First, we projected a sample subject's fMRI data onto the shared latent dimensions derived from cross-decomposition. For each dimension, we computed the temporal autocorrelation function:
\begin{equation}
R(\tau) = \frac{1}{T-\tau}\sum_{t=1}^{T-\tau}(x_t - \bar{x})(x_{t+\tau} - \bar{x})
\end{equation}
where $\tau$ represents the time lag in seconds, $T$ is the total number of timepoints, $x_t$ is the projected time series for a given dimension, and $\bar{x}$ is the mean of the time series. Dimensions were grouped by rank (1-10, 11-50, 51-100, 101-200), and mean autocorrelation functions were calculated for each group. To quantify characteristic timescales, we fitted an exponential decay function to each dimension's autocorrelation:
\begin{equation}
R_{fit}(\tau) = a \cdot e^{-\tau/\tau_0}
\end{equation}
where $\tau_0$ represents the characteristic decay time constant. 

\subsubsection{Reliable individual differences}
\paragraph{Individual difference matrices}
To assess whether there is reliable individual \textit{variability} along the ranks of these high-dimensional spectra, we constructed individual differences matrices (IDMs) that capture the pairwise similarities of subjects \cite{feilong2018reliable}. For each logarithmic bin $\kappa$ of latent dimensions, each cell of an IDM was computed from a pair of subjects $i$ and $j$ as:
\begin{equation}
     IDM_{c}(i, j) = \Sigma_{test,c}^{(i, j)}
\end{equation}
where $\Sigma_{test,c}^{(i, j)}$ represents the cross-validated covariance for the bin centered at $c$. These IDMs reflect individual differences in representations across segments of latent dimensions. Covariance values in IDMs displayed in Fig. \ref{fig:2} and \ref{fig:3} were normalized to z-scores for visualization only and statistical analyses were performed on raw covariance values.

After regressing out head motion confounds (intersubject framewise-displacement correlations and median framewise-displacement similarities) from \cite{savasegal}, we assessed reliability by computing Spearman correlations between mean IDMs derived from even and odd movies:
% \begin{equation}
%      \rho_{c} = \text{corr}\left(IDM_{c}^{even}, IDM_{c}^{odd}\right)
% \end{equation}
\begin{equation}
     \rho_{c} = \rho (IDM_{c}^{even}, IDM_{c}^{odd})
\end{equation}

\added{To ensure full independence between training and test data, we used a split-half approach where singular vectors were learned from even movies and applied separately to each odd movie to compute their IDMs. Reliability was assessed by correlating IDMs from different odd movies. We repeated this with train and test folds reversed and reported the average of the correlations across both folds. For the whole-brain analysis, we reported results from the even-fold test set only due to computational constraints.}

\paragraph{Controlling for voxelwise structure}
To distinguish fine-grained multivariate patterns from coarse-grained global effects, we constructed voxelwise inter-subject correlation (ISC) matrices. For each pair of subjects, we computed the mean Pearson correlation across all voxels in their response timeseries:
% \begin{equation}
% IDM_{ISC}(i, j) = \frac{1}{v}\sum_{k=1}^{v}\text{corr}(X_{test,k}, Y_{test,k})  
% \end{equation}
\begin{equation}
IDM_{ISC}(i, j) = \frac{1}{v}\sum_{k=1}^{v}\text{r}(X_{test,k}, Y_{test,k})  
\end{equation}
where $X_{test,k}$ and $Y_{test,k}$ represent the response timeseries for voxel $k$ in subjects $i$ and $j$, respectively. We then assessed whether the fine-grained patterns captured by our dimensional approach contained information beyond these coarse-scale similarities by computing partial correlations:
% \begin{equation}
% \rho_{c}' = \text{partial-corr}(IDM_{c}^{even}, IDM_{c}^{odd} \mid IDM_{ISC}^{even}, IDM_{ISC}^{odd})
% \end{equation}
\begin{equation}
\rho_{c}' = \rho (IDM_{c}^{even}, IDM_{c}^{odd} \mid IDM_{ISC}^{even}, IDM_{ISC}^{odd})
\end{equation}
where the partial correlation was computed by applying Spearman correlation to the residuals after regressing out the ISC matrices.

\paragraph{Bootstrapped confidence intervals}
To estimate confidence intervals, we implemented a bootstrap procedure with 1,000 resamples. For each resample, we randomly selected 90\% of subjects with replacement and recomputed IDM correlations, excluding self-similarity values \cite{kriegeskorte2008representational}. This generated a correlation distribution for each rank bin, from which we derived 95\% confidence intervals. We computed IDM correlations both before and after partialing out ISC effects, with FDR correction applied when comparing differences between these conditions.

\paragraph{Controlling for lower-rank dimensions}
To assess whether higher-rank dimensions contain unique reliable information beyond what is captured by lower-rank dimensions, we performed successive partial correlation analyses. For each dimensional bin $c > 1$, we sequentially regressed out IDMs from bins $1, ..., c-1$ from both the even- and odd-movie IDMs at bin $c$ in an iterative manner. We then computed the Spearman correlation between the final residual even- and odd-movie IDMs. Head motion confounds were regressed out prior to these analyses. Statistical significance was assessed using the same bootstrap procedure described above (1,000 resamples, 95\% confidence intervals, Bonferroni correction).

\paragraph{Dimensional specificity}
To examine whether different ranges of latent dimensions capture distinct patterns of individual differences, we computed correlations between IDMs from different dimensional ranges. For each dimension bin $c$, we computed correlations between its even-movie IDM and odd-movie IDMs from all ranges. We then quantified the specificity of individual differences by comparing the diagonal elements (same-range correlations) to the mean of off-diagonal elements (different-range correlations) in each row. The diagonal dominance was computed by subtracting different-range correlations from the same-range correlations at each range.

To examine regional specificity of individual differences, we extended our IDM correlation approach across regions. We constructed a correlation matrix examining the relationships between IDMs from all combinations of dimensional ranges and cortical regions. Specifically, for each region pair (early visual, ventral temporal, lateral temporal, and posterior parietal-cingulate) and each dimensional range pair, we computed Spearman correlations between even-movie IDMs from one region-range combination and odd-movie IDMs from another. We conducted this analysis both with original IDMs and after controlling for coarse-scale effects through the partial correlation procedure described above. 

\subsubsection{Neural-behavioral similarity analysis}

To examine relationships between neural individual differences and subjective movie interpretations, we analyzed the free recall data from Sava-Segal et al.\cite{savasegal} Following the original study, we used transcribed verbal descriptions collected after each movie viewing, and we encoded these descriptions into a 512-dimensional semantic space using Google's Universal Sentence Encoder \cite{cer2018universal}. We constructed behavioral similarity matrices by computing pairwise cosine similarities between participants' semantic embeddings. \added{Seva-Segal et al.\cite{savasegal} selected USE because it was specifically trained to identify similarities between sentence pairs and because it best differentiated recall content between movies compared to alternative sentence encoders (BERT BASE, MiniLM, and MPNet). We repeated our key analyses using these alternative encoders and confirmed that the results were qualitatively similar.}  

We examined whether pairs of subjects with similar neural representational structure showed similar movie interpretations. First, we regressed out head motion confounds from all matrices. Then, we computed Spearman correlations between the behavioral similarity matrices and neural IDMs across different ranges of latent dimensions. This analysis was performed both with the original neural IDMs and with IDMs after controlling for voxelwise ISC matrices through partial correlation, allowing us to assess whether fine-grained neural patterns specifically contributed to behavioral similarities.

Statistical significance was assessed through permutation tests with 1,000 iterations. For each iteration, subject indices were randomly shuffled to generate a null distribution of neural-behavioral correlations. P-values were computed as the proportion of permuted correlations exceeding the observed correlation and were FDR-corrected across dimensional ranges.

\paragraph{Controlling for other dimensional ranges}
\added{To assess whether different dimensional ranges contain unique information about behavioral similarity beyond other ranges, we performed successive semi-partial correlation analyses. For each of the three dimensional ranges (1-10, 11-100, and 101-1000), we iteratively regressed out the neural IDMs from the other two dimensional ranges from the neural IDM of interest. Specifically, for a given dimensional range, we sequentially removed the influence of the other two ranges: we first regressed out the neural IDM from one of the other ranges to obtain a residual neural IDM, then regressed out the neural IDM from the remaining range from this residual. We then computed the Spearman correlation between the final residual neural IDM and the behavioral IDM. Head motion confounds were regressed out from all neural IDMs prior to these analyses. We also performed this analysis after controlling for voxelwise ISC matrices (as described above). The same permutation testing procedure (1,000 iterations, FDR correction) was applied to assess statistical significance.}

\subsubsection{Concreteness and shared neural covariance}
\paragraph{Mean covariance analysis}
\added{To relate the recall narrative concreteness to neural patterns across dimensional bins (ranks 1-10, 11-100, 101-1000), we computed each subject's mean neural covariance for each dimensional bin by averaging all IDM values involving that subject (41 pairwise comparisons per subject given n=42 subjects). We then computed the Spearman correlations between these subject-level neural values and the recall narrative concreteness scores (See "Narrative concreteness"). Statistical significance was assessed through permutation tests with 1,000 iterations, FDR-corrected. To isolate unique variance per bin, we additionally performed successive semi-partial correlations controlling for the other two bins, iteratively removing their effects before computing correlations with concreteness scores (Fig. \ref{figS6}).}

\paragraph{Dimensionality count analysis}
\added{We quantified individual neural dimensionality by counting, for each subject, how many dimensions showed significant covariance (permutation test, 99.5th percentile threshold) across all subject pairs involving that subject, evaluating the significance of each dimension without binning. We then computed the subject-level dimensionality counts averaged across movies and computed the Spearman correlation with the recall concreteness scores.}

\subsection{Data availability} The naturalistic audio-visual movie fMRI dataset is available at: \url{https://openneuro.org/datasets/ds004516}.

\subsection{Code availability} All code for analyses are available at: \url{https://github.com/kelseyhan-jhu/idiosyncratic-neural-geometry}.

\newpage

%%%%%%%%%%%%%%%% START OF SUPPLEMENT %%%%%%%%%%%%%%%

% Figures, tables, equations and pages in the supplement are numbered S1, S2 etc.
\renewcommand{\thefigure}{S\arabic{figure}}
\renewcommand{\thetable}{S\arabic{table}}
\renewcommand{\theequation}{S\arabic{equation}}
\setcounter{figure}{0}
\setcounter{table}{0}
\setcounter{equation}{0}
% \setcounter{page}{1} % not 0 as \newpage already started a supplementary page
% References continue the numbering from the main text.

% %%%%%%%%%%%%%%%% SUPPLEMENT TITLE PAGE %%%%%%%%%%%%%%%

% \begin{center}
% \section*{Supplementary Materials for\\ \scititle}
% Chihye Han,
% Michael F. Bonner$^{\ast}$\\ 
% \small$^\ast$Corresponding author. Email: mfbonner@jhu.edu\\
% \end{center}

% \subsubsection*{This PDF file includes:}
% % Methods\\
% Figures S1 to S3\\

\newpage

\begin{figure} 
	\centering
	\includegraphics[width=1\textwidth]{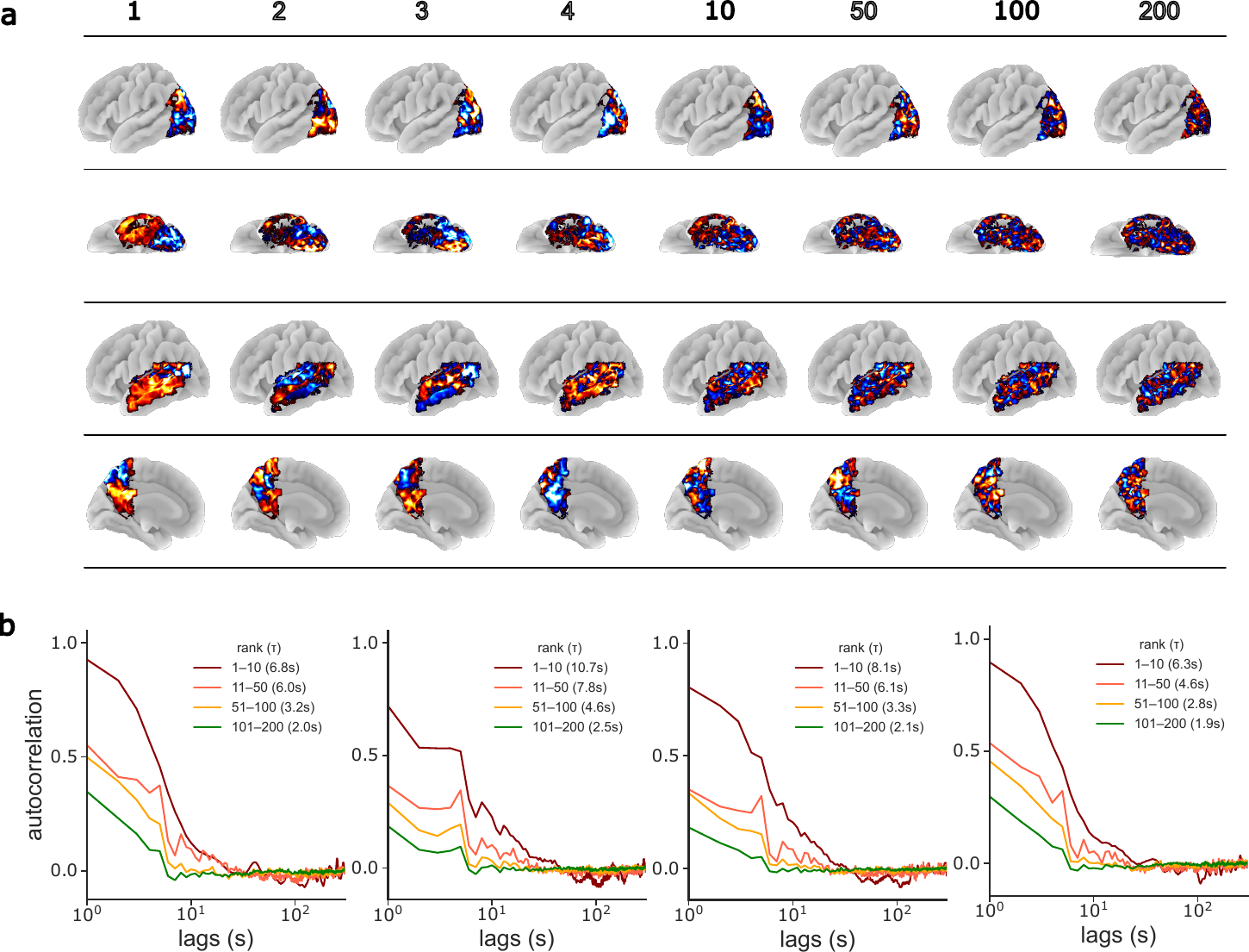} 
	\caption{\textbf{Latent dimensions reveal systematic spatiotemporal organization from coarse to fine patterns.} (\textbf{a}) Visualization of latent component loadings across dimension ranks (numbers above each brain) demonstrates a hierarchical organization of neural activity patterns. Low-rank dimensions (1-10) capture large-scale spatial patterns with high variance, while higher-rank dimensions progressively represent finer-grained spatial structure. From top to bottom: early visual, ventral temporal, lateral temporal, and posterior parietal-cingulate  regions. (\textbf{b}) Temporal autocorrelation analysis quantifies the timescale of neural dynamics across dimension ranks. Plots show autocorrelation values as a function of time lag (in seconds) for different dimensional ranges. From left to right: early visual, ventral temporal, lateral temporal, and posterior parietal-cingulate regions. Higher-rank dimensions (green lines) show faster temporal fluctuations with shorter autocorrelation timescales compared to lower-rank dimensions (red and brown lines), revealing a systematic relationship between spatial and temporal scales across the dimensional hierarchy within each cortical region.
        }
	\label{figS2} 
\end{figure}

% \begin{figure} % Do not use \begin{figure*}
% 	\centering
% 	\includegraphics[width=1\textwidth]{contribution.pdf} % for an image file named example_figure.*
% 	% Pick an appriopriate width for the size of the image

% 	% Captions go below figures
% 	\caption{how much of isc reliability is explained by each dimensional covariance structure
%         }
% 	\label{figS3} % give each figure a logical label name
% \end{figure}

\begin{figure} 
	\centering
	\includegraphics[width=1\textwidth]{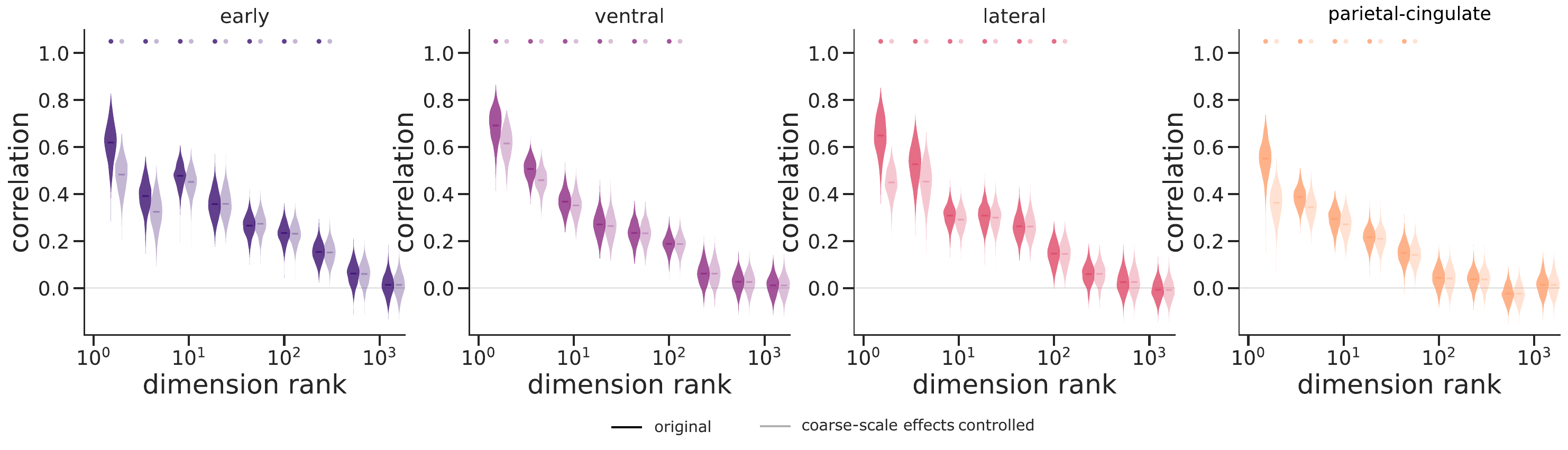} 
	\caption{\added{\textbf{High-dimensional individual differences remain reliable after controlling for low-dimensional structure.} Split-half reliability of IDMs observed across dimensional ranges in Fig. \ref{fig:2}\textbf{B} persists even after sequentially removing contributions from lower-dimensional ranges. Dark shades represent correlations between original IDMs; lighter shades show correlations after controlling for coarse-scale effects through partialing out voxelwise ISC IDMs. Grey horizontal line indicates zero correlation. Dots indicate statistical significance ($p < 0.05$, Bonferroni-corrected); bootstrap confidence intervals from 1,000 resamples. From left to right: early visual, ventral temporal, lateral temporal, and posterior parietal-cingulate regions.}}
	\label{figS3} 
\end{figure}

\begin{figure} 
	\centering
	\includegraphics[width=1\textwidth]{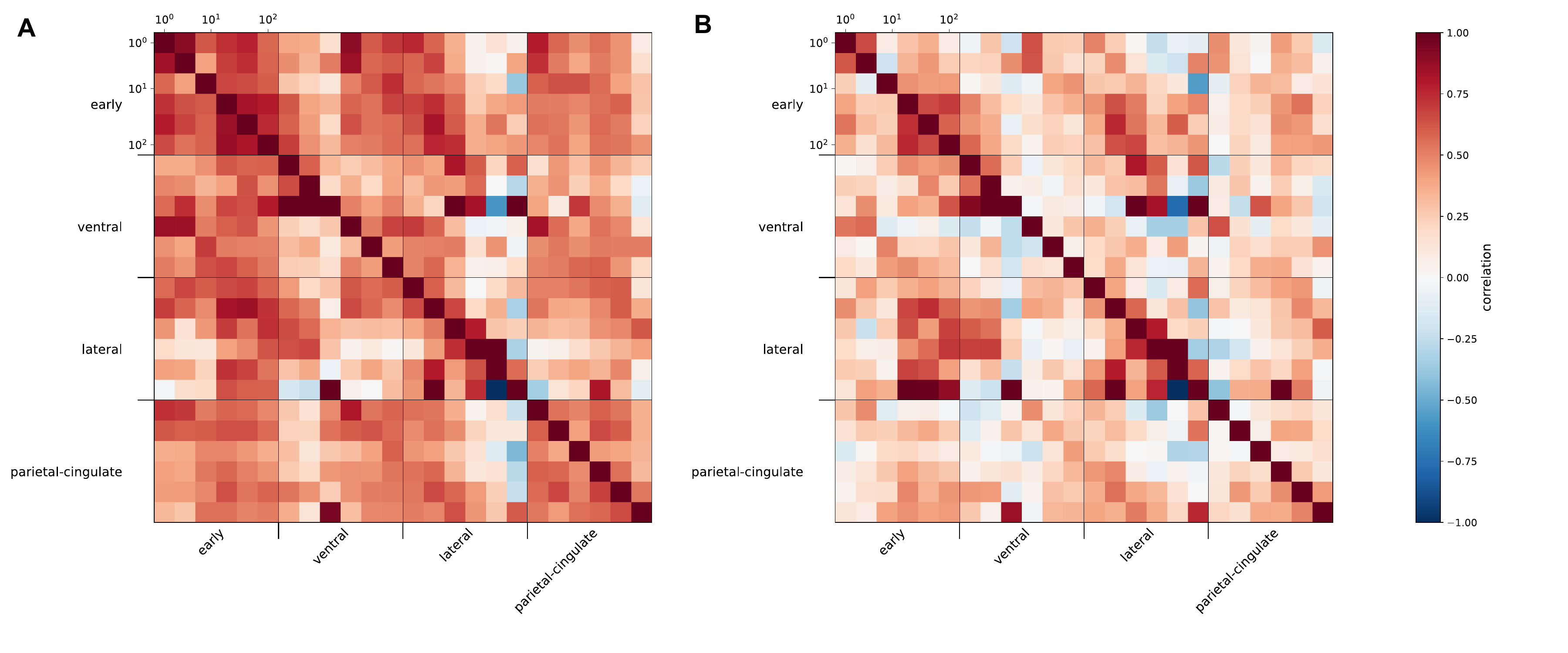} 
	\caption{\textbf{Region-specific patterns of individual differences emerge at each dimensional range.} (\textbf{a}) Correlation matrices show relationships between individual differences matrices (IDMs) across regions and dimensional ranges. Each element represents the correlation between even-movie IDMs from one range in a region (rows) and odd-movie IDMs from another range in a region (columns), \added{normalized by the corresponding diagonal reliabilities such that diagonal values equal 1}. The log-scaled axis labels indicate dimensional ranges. The diagonal blocks show within-region correlations across matching dimensional ranges, while off-diagonal blocks show cross-region correlations.
    (\textbf{b}) Same analysis after controlling for coarse-scale effects by partialing out voxelwise inter-subject correlations. The more pronounced diagonal structure demonstrates that removing coarse-grained patterns reveals region-specific patterns of individual variability that are unique to each cortical area and dimensional range. This analysis complements Fig. \ref{fig:3} by showing that individual differences not only vary across dimensions within regions but also exhibit distinct patterns across regions at corresponding dimensional ranges.
    }
	\label{figS4} 
\end{figure}

\begin{figure} 
	\centering
	\includegraphics[width=1\textwidth]{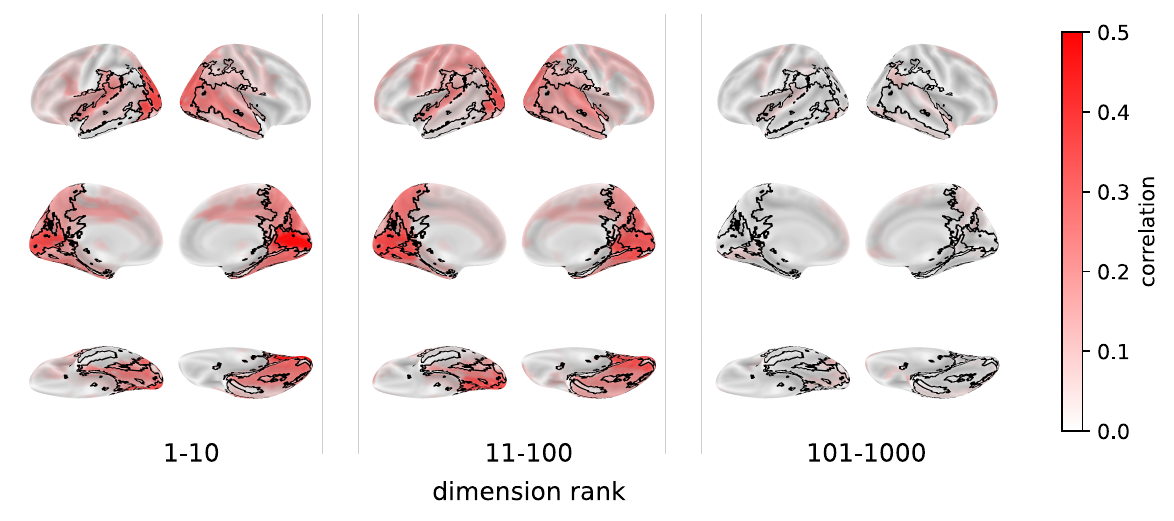} 
	\caption{\added{\textbf{High-dimensional individual differences are strongest in sensory and semantic cortices.} (\textbf{A}) Maps show correlation across fine-grained brain parcellations for dimensions 1-10 (left), 11-100 (middle), and 101-1000 (right). Individual differences extending to high-dimensional ranges are strongest in sensory regions (visual and auditory cortices) and semantic/default mode regions (precuneus, angular gyrus), modest in memory-related regions (hippocampus, medial prefrontal cortex), and diminished in other regions. This spatial distribution indicates that multidimensional individual differences reflect meaningful stimulus-driven variability concentrated in regions engaged by naturalistic audiovisual content. Maps show raw correlations between odd-movie IDMs without thresholding. Black outlines indicate main analysis ROIs.}}
	\label{figS5} 
\end{figure}

\begin{figure} 
	\centering
	\includegraphics[width=1\textwidth]{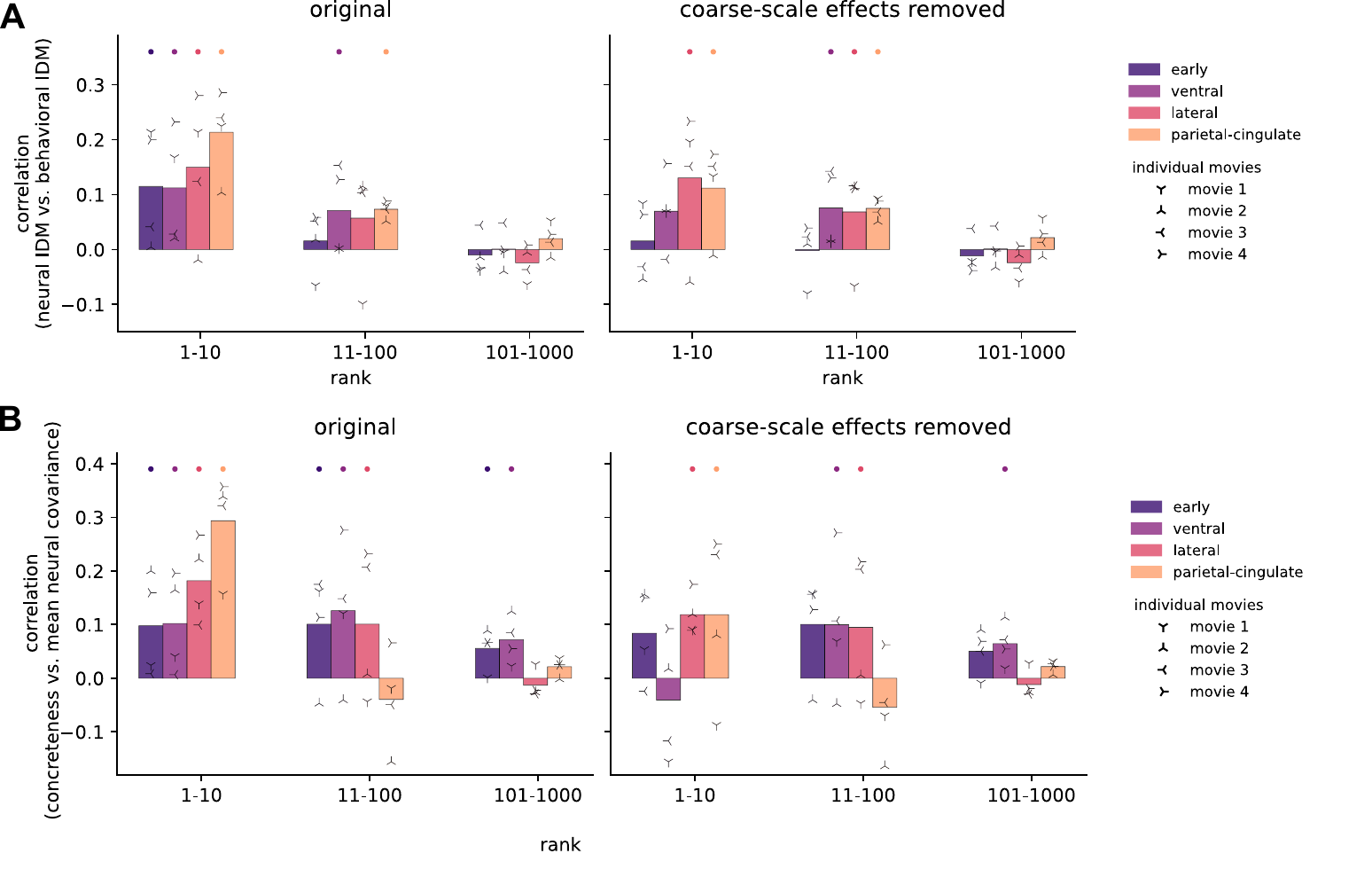} 
	\caption{\added{\textbf{Unique contributions of different dimensional ranges to behavioral predictions.} (\textbf{A}) Partial correlations between neural IDMs and behavioral similarity, controlling for all other dimensional ranges, to supplement Fig. 4\textbf{B}. (\textbf{B}) Partial correlations between shared neural variance and narrative concreteness, controlling for all other dimensional ranges, to supplement Fig. 5\textbf{B}. Left shows original neural covariance; Right shows results after removing coarse-scale effects. Bars show mean partial correlations; dots indicate statistical significance ($p < 0.05$, FDR-corrected). Small symbols represent individual movies. While attenuated compared to zero-order correlations, several dimensional ranges maintained significant unique contributions.}}

	\label{figS6} 
\end{figure}

% \begin{figure} 
% 	\centering
% 	\includegraphics[width=1\textwidth]{figS6_REV.pdf} 
% 	\caption{\textbf{Test.} (\textbf{a}) To Do
%     }
% 	\label{figS6} 
% \end{figure} 

% \begin{figure} 
% 	\centering
% 	\includegraphics[width=1\textwidth]{figS7_REV.pdf} 
% 	\caption{\textbf{Test.} (\textbf{a}) To Do
%     }
% 	\label{figS7} 
% \end{figure} 

% \begin{figure} 
% 	\centering
% 	\includegraphics[width=1\textwidth]{figS6_REV.pdf} 
% 	\caption{\textbf{Region-specific patterns of individual differences emerge at each dimensional range.} (\textbf{a}) To Do
%     }
% 	\label{figS6} 
% \end{figure} 

%%%%%%%%%%%%%%%% REFERENCES %%%%%%%%%%%%%%%

\clearpage
\bibliography{science_template} % for a file named science_template.bib
\bibliographystyle{sciencemag}

% %%%%%%%%%%%%%%%% SUPPLEMENTARY TABLES %%%%%%%%%%%%%%%

% %%%%%%%%%%% CAPTIONS FOR OTHER SUPPLEMENTARY FILES %%%%%%%%%%

%%%%%%%%%%%%%%%% SUPPLEMENTARY REFERENCES %%%%%%%%%%%%%%%

\end{document}